\documentclass[epj]{svjour}
\usepackage{graphics}
\begin{document}
\title{Proxy-SU(3) symmetry in heavy nuclei: Foundations}

\author{Dennis Bonatsos\inst{1} \and \underline{I. E. Assimakis}\inst{1} \and N. Minkov\inst{2} \and Andriana Martinou\inst{1} \and R. B. Cakirli\inst{3} \and R. F. Casten\inst{4,5} \and
K. Blaum\inst{6}} 
%
%
\institute{Institute of Nuclear and Particle Physics, National Centre for Scientific Research 
``Demokritos'', GR-15310 Aghia Paraskevi, Attiki, Greece
\and Institute of Nuclear Research and Nuclear Energy, Bulgarian Academy of Sciences, 72 Tzarigrad Road, 1784 Sofia, Bulgaria
\and Department of Physics, University of Istanbul, Istanbul, Turkey
\and Wright Laboratory, Yale University, New Haven, Connecticut 06520, USA
\and Facility for Rare Isotope Beams, 640 South Shaw Lane, Michigan State University, East Lansing, MI 48824 USA
\and  Max-Planck-Institut f\"{u}r Kernphysik, Saupfercheckweg 1, D-69117 Heidelberg, Germany}

\date{Received: date / Revised version: date}
%
\abstract{
An approximate SU(3) symmetry appears in heavy deformed even-even nuclei, by omitting the intruder Nilsson orbital of highest total
angular momentum and replacing the rest of the intruder orbitals by the orbitals which have escaped to the next lower major shell. 
The approximation is based on the fact that there is a one-to-one correspondence between the orbitals of the two sets, based on pairs 
of orbitals having identical quantum numbers of orbital angular momentum, spin, and total angular momentum. The accuracy of the approximation 
is tested through calculations in the framework of the Nilsson model in the asymptotic limit of large deformations, focusing attention on the changes 
in selection rules and in avoided crossings caused by the opposite parity of the proxies with respect to the substituted  orbitals. 
} 
\maketitle
\section{Introduction}
\label{intro}

The proxy-SU(3) scheme \cite{PRC1,PRC2}, already described in Ref.  \cite{IoBonat}, is a new approximate scheme applicable in medium-mass and heavy deformed nuclei, able to provide 
predictions for various nuclear properties, as it will be shown in Refs. 
\cite{IoMartinou,IoSarant}. In this contribution we are going to justify the relevant approximations 
by applying the proxy-SU(3) assumptions to the Nilsson model \cite{Nilsson1,Nilsson2}. 

\section{The Nilsson model}

The Nilsson Hamiltonian \cite{Nilsson1,Nilsson2} contains a harmonic oscillator with cylindrical symmetry, 
\begin{equation}
H_{osc}={ {\bf p}^2 \over 2M} +{1\over 2} M (\omega_z^2 z^2 + \omega^2_\perp (x^2+y^2)),
\end{equation}
where $M$ is the nuclear mass, ${\bf p}$ is the momentum, and  
the rotational frequencies $\omega_z$ and $\omega_\perp$ are related to the deformation parameter $\epsilon$ by
\begin{equation}
\omega_z = \omega_0 \left( 1 -{2\over 3} \epsilon \right), \qquad  \omega_\perp = \omega_0 \left( 1 +{1\over 3} \epsilon \right), 
\end{equation}
leading to 
\begin{equation}
\epsilon= {\omega_\perp - \omega_z \over \omega_0},
\end{equation}
with  $\epsilon > 0$ corresponding to prolate shapes and $\epsilon < 0$ corresponding to oblate shapes. In addition it contains a spin-orbit term and an angular momentum squared term, 
the total Hamiltonian having the form  
\begin{equation}
H=H_{osc} + v_{ls} \hbar \omega_0 ({\bf l} \cdot {\bf s}) + v_{ll} \hbar \omega_0 ({\bf l}^2 - \langle {\bf l}^2\rangle_N),
\label{Nil}\end{equation} 
where $\bf l$ is the angular momentum, ${\bf s}$ is the spin, 
\begin{equation}
\langle {\bf l}^2\rangle_N= {1\over 2} N(N+3)
\end{equation}
is the average of the square of the angular momentum $\bf l$ within the $N$th oscillator shell,
and  $v_{ls}$ and $v_{ll}$ are constants determined from the available data on intrinsic nuclear spectra \cite{BM}, their values being given in Table 5-1 of Ref. \cite{BM}. 
The eigenvalues of $H_{osc}$ are 
\begin{equation}\label{Hdiag}
E_{osc}= \hbar \omega_0 \left(N +{3\over 2} - {1\over 3} \epsilon (3n_z-N) \right). 
\end{equation}   

\section{Calculation of matrix elements}

In order to simplify the calculation of matrix elements, one can choose a more convenient basis.
Using the creation and annihilation operators $a^\dagger_x$, $a_x$, $a^\dagger_y$, $a_y$ for the quanta of the harmonic oscillator 
in the Cartesian coordinates $x$ and $y$, one can define creation and annihilation operators \cite{Nilsson2,MN}
\begin{equation}
R^+ ={1\over \sqrt{2}} (a_x^\dagger + i a_y^\dagger), \qquad R ={1\over \sqrt{2}} (a_x  - i a_y), 
\end{equation}
\begin{equation}
S^+ ={1\over \sqrt{2}} (a_x^\dagger - i a_y^\dagger), \qquad S ={1\over \sqrt{2}} (a_x +  i a_y).
\end{equation}
These operators satisfy the commutation relations 
\begin{equation}
[R, R^\dagger]=[S,S^\dagger]=1. 
\end{equation}
In this way one goes over to a new basis, $|n_z r s \Sigma \rangle $, where $r$  ($s$) is the number of quanta related to the harmonic oscillator
formed by $R^\dagger$ and $R$  ($S^\dagger$ and $S$).  The following relations hold 
\begin{equation}\label{rs}
n_\perp = r+s =N-n_z, \qquad \Lambda = r-s, 
\end{equation}
where $n_\perp$ is the number of quanta perpendicular to the $z$-axis.  

It is then a straightforward task, described in detail in Ref. \cite{Nilsson2}, to calculate the matrix elements of the ${\bf l} \cdot {\bf s}$ and ${\bf l}^2$ operators in the new basis.

The spin-orbit term, ${\bf l} \cdot {\bf s}$, has diagonal matrix elements 
\begin{equation}
\langle n_z r s \Sigma |{\bf l} \cdot {\bf s} |  n_z r s \Sigma\rangle = (r-s) \Sigma = \Lambda \Sigma, 
\end{equation}
as well as non-diagonal matrix elements 
\begin{equation}
\langle n_z-1, r+1, s, \Sigma-1 |{\bf l} \cdot {\bf s} |  n_z r s \Sigma\rangle = -{1\over \sqrt{2}} \sqrt{n_z (r+1)},
\end{equation}
\begin{equation}
\langle n_z+1, r, s-1, \Sigma-1 |{\bf l} \cdot {\bf s} |  n_z r s \Sigma\rangle = {1\over \sqrt{2}} \sqrt{(n_z+1) s},
\end{equation}
\begin{equation}
\langle n_z+1, r-1, s, \Sigma+1 |{\bf l} \cdot {\bf s} |  n_z r s \Sigma\rangle = -{1\over \sqrt{2}} \sqrt{(n_z+1) r}, 
\end{equation}
\begin{equation}
\langle n_z-1, r, s+1, \Sigma+1 |{\bf l} \cdot {\bf s} |  n_z r s \Sigma\rangle = {1\over \sqrt{2}} \sqrt{n_z (s+1)}. 
\end{equation}

The orbital angular momentum term, ${\bf l}^2$, has diagonal matrix elements 
\begin{equation}\label{l2diag}
\langle n_z r s \Sigma |{\bf l}^2 |  n_z r s \Sigma\rangle = 2 n_z(r+s+1) +(r+s)+(r-s)^2, 
\end{equation}
as well as non-diagonal matrix elements

$$\langle n_z+2, r-1, s-1, \Sigma |{\bf l}^2 |  n_z r s \Sigma\rangle = $$
\begin{equation} 
 -2\sqrt{(n_z+2)(n_z+1) r s }, 
\end{equation}

$$ \langle n_z-2, r+1, s+1, \Sigma |{\bf l}^2 |  n_z r s \Sigma\rangle = $$ 
\begin{equation}
-2\sqrt{(n_z-1) n_z (r+1)(s+1)}.
\end{equation} 

\section{Numerical results}

Numerical results for the matrix elements  of ${\bf l} \cdot {\bf s}$ for the 126-184 and sdgi shells are given in Table I, and those for the matrix elements of ${\bf l}^2$ for the same shells are given in Table II. Furthermore, numerical results for the full Hamiltonian for $\epsilon =0.3$ for the 126-184 and sdgi shells are given in Table III, while Nilsson-like diagrams involving either just the diagonal terms of the Hamiltonian, or obtained through the diagonalization of the full Hamiltonian, are plotted for the same shells in Fig.~1. Many comments are in place.

In each of the Tables I, II, III, the upper table corresponds to the original Nilsson model 126-184 shell, 
while the lower table represents the proxy-SU(3) approximation to it, which is an sdgi shell.  
The lower table has one row and one column less than the upper table, since the orbital 
with the highest angular momentum in the 126-184 shell, 15/2[707], has no counterpart 
in the proxy-SU(3) sdgi shell. All tables are symmetric, thus only the upper half is shown. 

All tables are divided into four blocks. 
The upper left block involves matrix elements among the normal parity orbitals, i.e. the orbitals which belong to the sdgi shell and remain in the 126-184 shell after the defection of the 1i$_{13/2}$ orbital to the major shell below. Therefore the upper left blocks of the 126-184 and sdgi tables 
are identical.  

The lower right block involves matrix elements among the abnormal orbitals, members of the 
1j$_{15/2}$ orbital which has invaded the sdgi shell coming from the major shell above. 
The 126-184 and sdgi results are either identical, or very similar, due to the fact that 
during the proxy-SU(3) approximation all angular momentum projections (orbital angular 
momentum, spin, total angular momentum) remain unchanged.  

The upper right block of the 126-184 tables contains only vanishing matrix elements, 
since it connencts levels of opposite parity, namely positive parity sdgi orbitals 
to negative parity 1j$_{15/2}$ orbitals. In the upper right block of the sdgi tables, though,
a few non-vanishing matrix elements appear, connecting the original positive parity sdgi orbitals 
to the proxies of the 1j$_{15/2}$ orbitals, which are positive parity 1i$_{13/2}$ orbitals.  
These extra non-vanishing matrix elements represent the ``damage'' caused by the proxy-SU(3) 
approximation, i.e., by the replacement of the 1j$_{15/2}$ orbitals (except the 15/2[707] one)
by their 1i$_{13/2}$ proxies. 

The size of these extra non-vanishing matrix elements is comparable to the magnitude 
of the diagonal matrix elements in the case of the ${\bf l} \cdot {\bf s}$ and ${\bf l}^2$ operators,
as one can see in Tables I and II.
However, as one can see in Table III, the extra non-vanishing matrix elements are  
much smaller (by at least one order of magnitude) than the diagonal matrix elements.
The reason behind this difference is the small values of the $v_{ls}$ 
and $v_{ll}$ coefficients ($-0.127$ and $-0.0206$ respectively, as given in Table 5-1 
of Ref. \cite{BM}), by which the matrix elements of  ${\bf l} \cdot {\bf s}$ and ${\bf l}^2$ 
are multiplied before entering Table III. 

The smallness of the extra non-vanishing matrix elements means that the diagonalization 
of the Nilsson Hamiltonian for various values of the deformation $\epsilon$ will yield
very similar results for the 126-184 and proxy-SU(3) sdgi shells, as seen in Fig. 1,
indicating that the Nilsson diagrams are very little affected by these extra non-vanishing matrix 
elements and thus proving that the proxy-SU(3) approximation is a good one. 

In a similar way one can see that good results are also obtained for the 
28-50, 50-82, and 82-126 shells in comparison to their pf, sdg, and pfh proxy-SU(3) 
counterparts \cite{PRC1}. 

\section{Conclusions} 

By applying the proxy-SU(3) assumptions to the Nilsson model, we prove that the Nilsson 
diagrams of the 126-184 shell and of its proxy-SU(3) sdgi counterpart are very similar, 
thus justifying the use of the proxy-SU(3) scheme, some applications of which for the calculation 
of nuclear properties will be discussed in Refs. \cite{IoMartinou,IoSarant}.

\begin{figure*}[htb]

\resizebox{0.99\textwidth}{!}{%
  \includegraphics{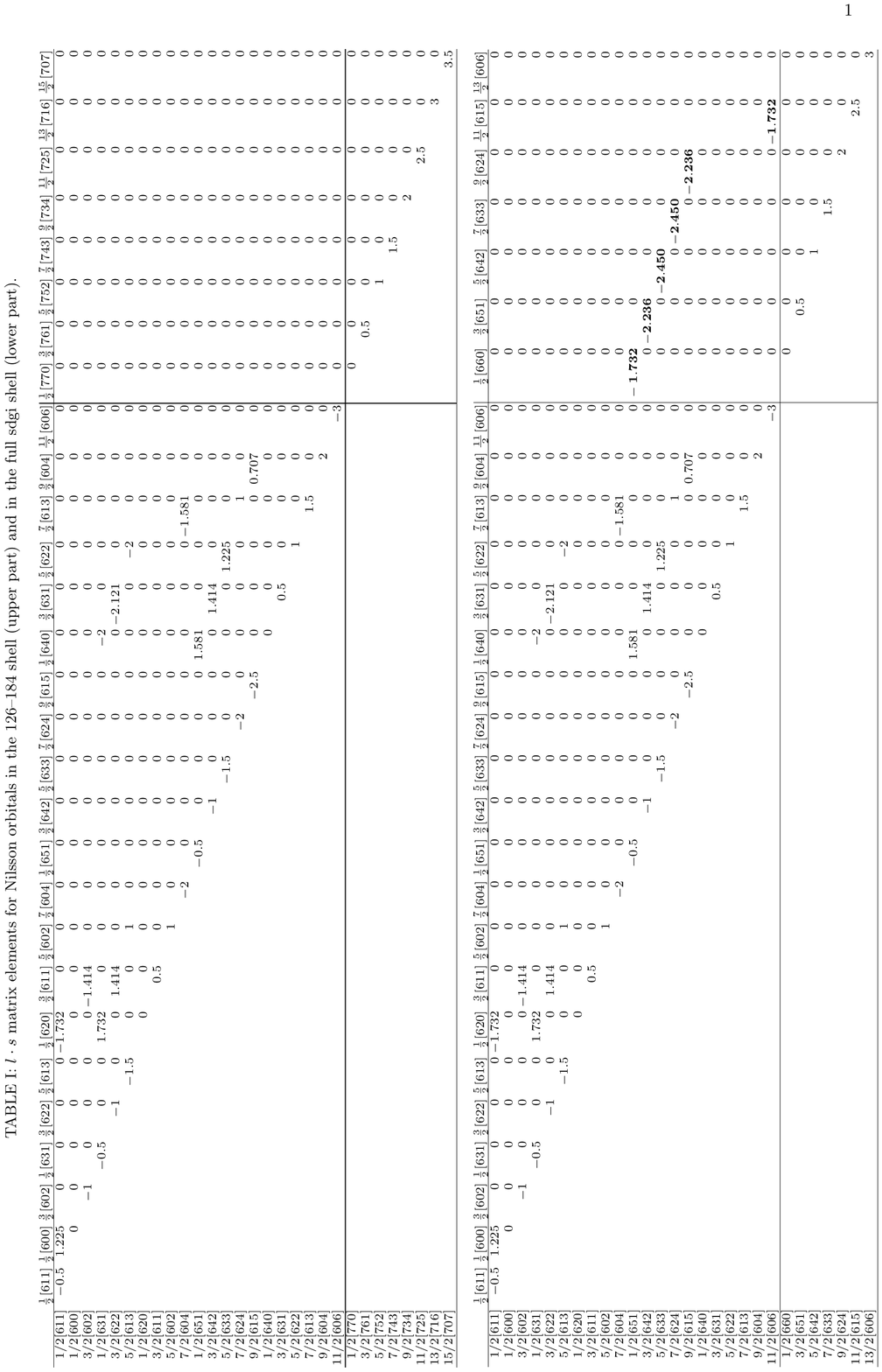}
}

\end{figure*}

\begin{figure*}[htb]

\resizebox{0.99\textwidth}{!}{%
  \includegraphics{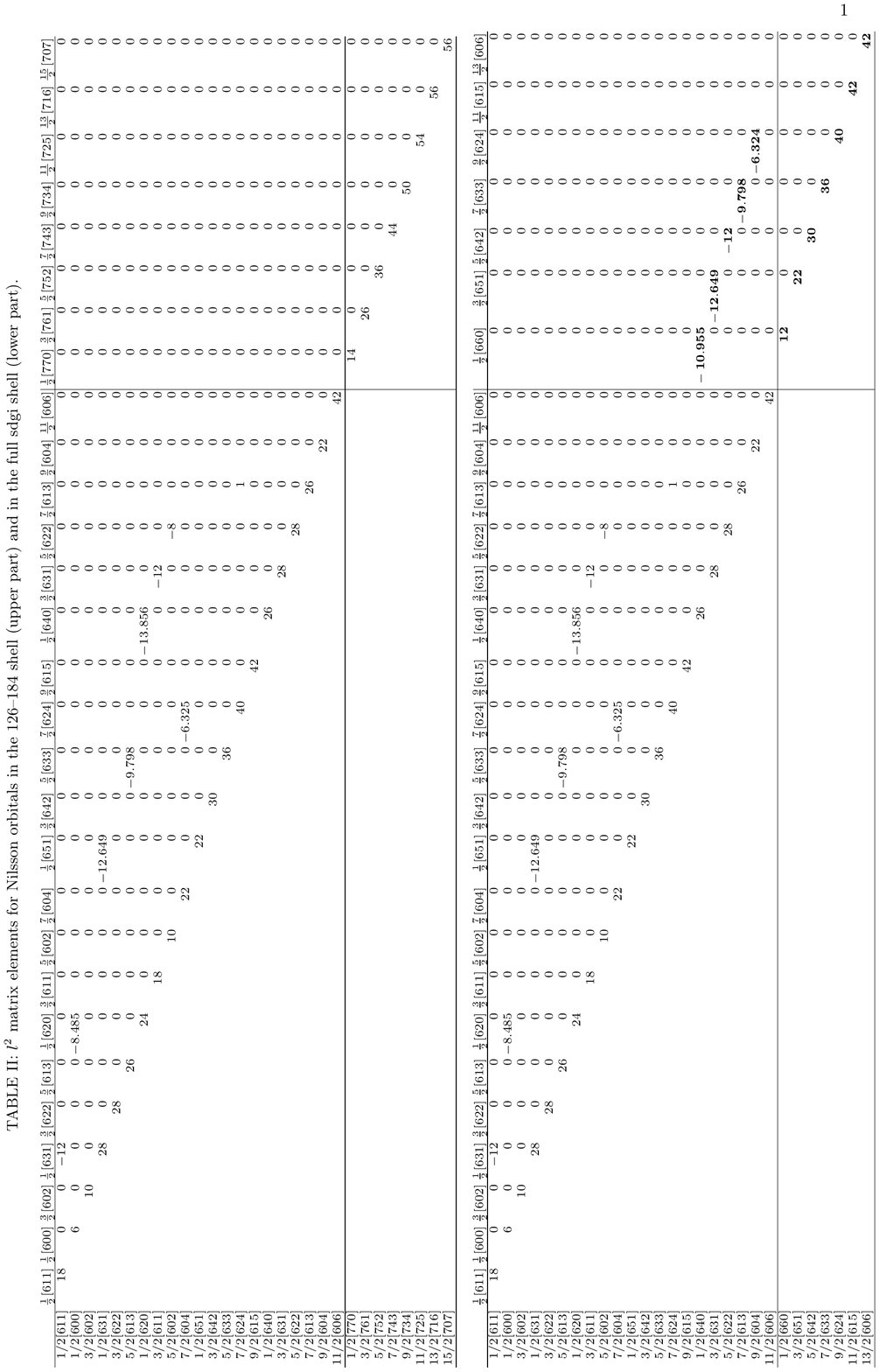}
}

\end{figure*}

\begin{figure*}[htb]

\resizebox{0.99\textwidth}{!}{%
  \includegraphics{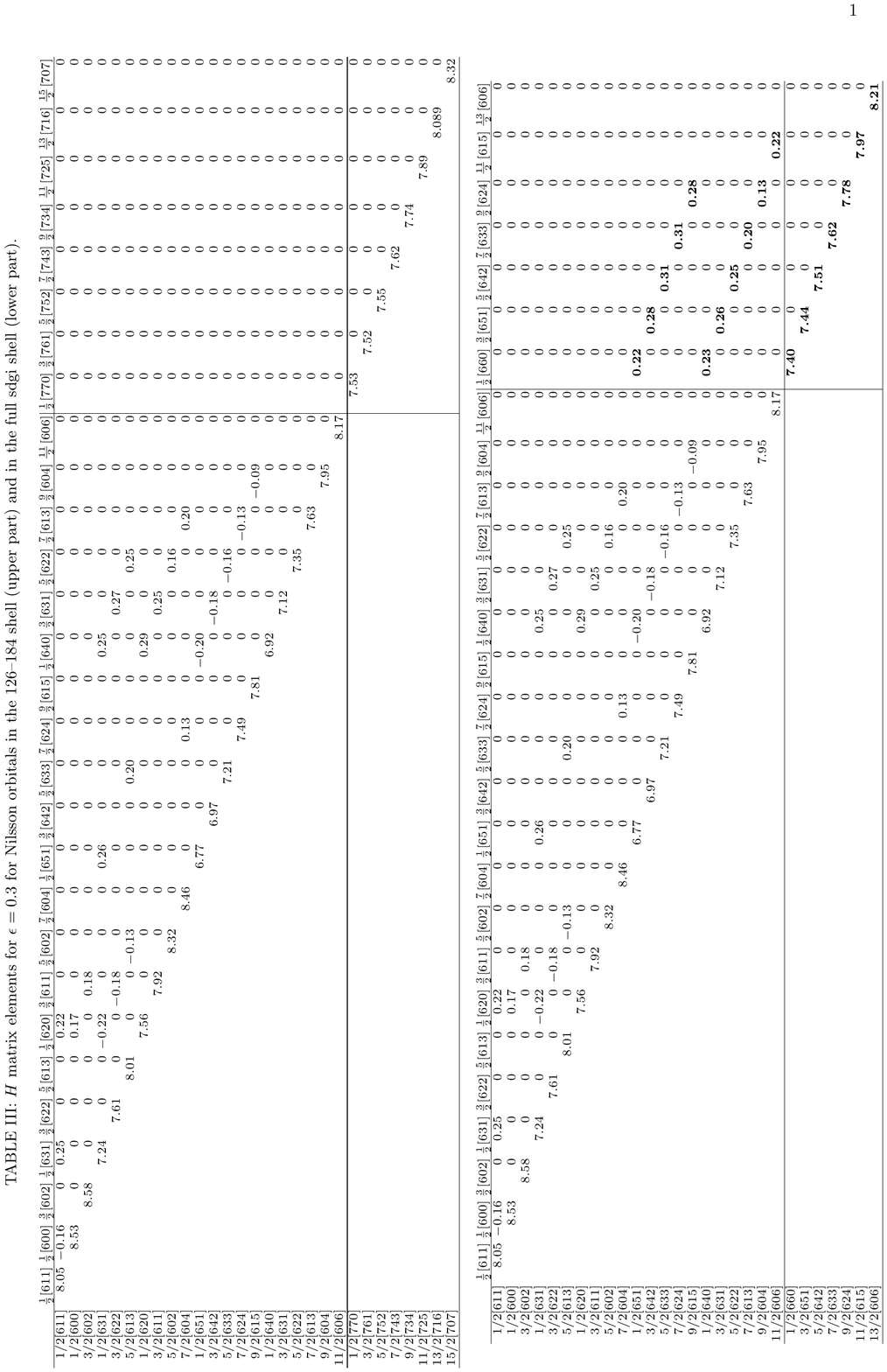}
}

\end{figure*}

\begin{figure*}[htb]

\resizebox{0.90\textwidth}{!}{%
  \includegraphics{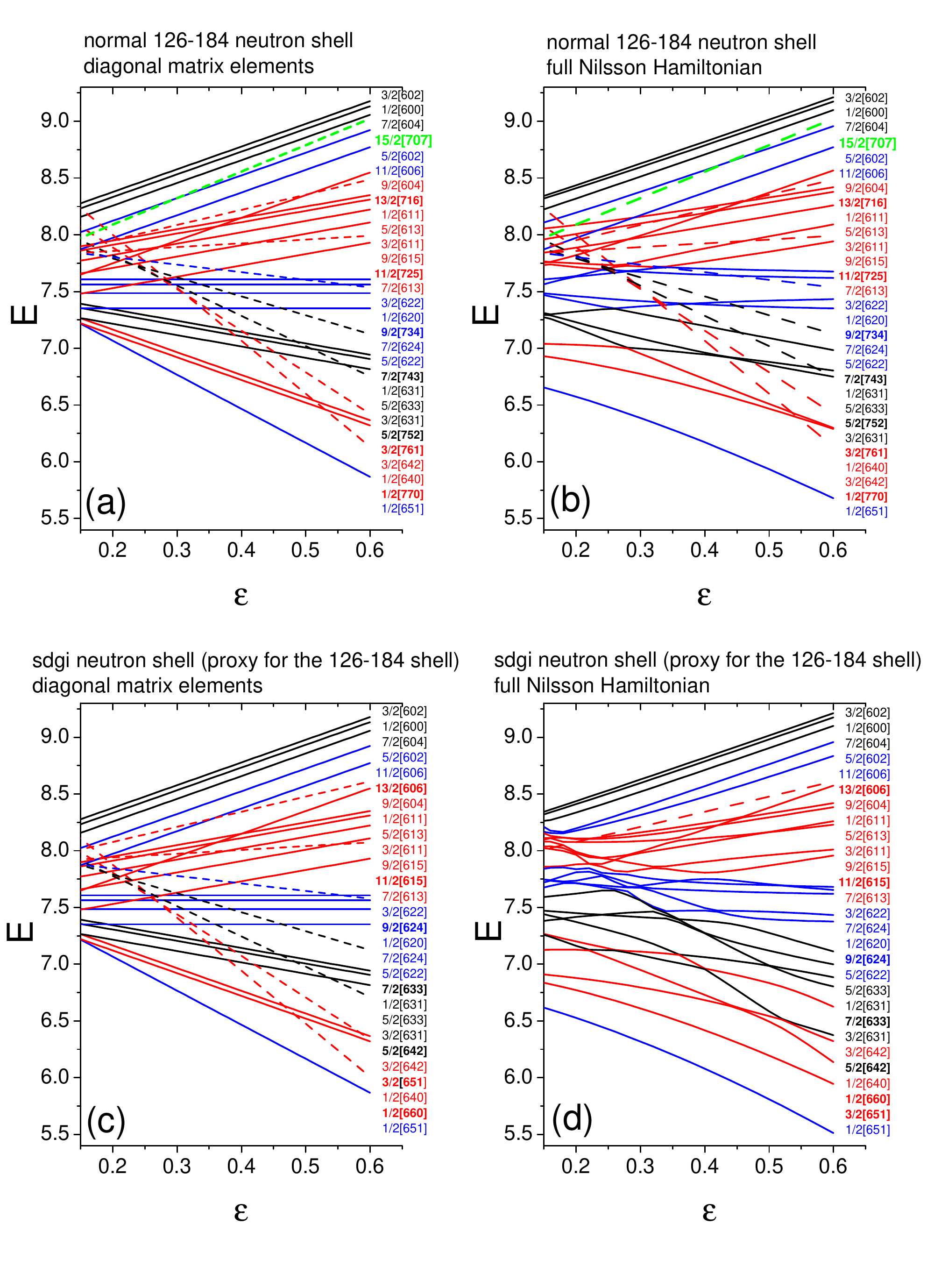}
}

\caption{Diagonal matrix elements (in units of $\hbar \omega_0$) of the Nilsson Hamiltonian for the 126-184 (a) and sdgi (c) neutron shells compared 
to the results of the full diagonalization for the 126-184 (b) and sdgi (d) neutron shells, as functions of the deformation parameter $\epsilon$. 
The Nilsson parameters are taken from Table 5-1 of Ref. \cite{BM}. The 1j$_{15/2}$ orbitals in (a) and (b), as well as the 1i$_{13/2}$ orbitals in (c), are 
indicated by dashed lines. The 1j$_{15/2}$ orbital labels in (a) and (b), as well as the 1i$_{13/2}$ orbital labels in (c) and (d), appear in boldface. Orbitals are grouped in color only to facilitate visualizing 
the patterns of orbital evolution. Note that the Nilsson labels at the right are always in the same order
as the energies of the orbitals as they appear at the right as well (largest deformation shown). 
Therefore, in some cases the order of the Nilsson orbitals changes slightly from panel to panel.
 } 
\end{figure*}

\end{document}